\documentclass[final,times,twocolumn]{elsarticle}
\usepackage{colordvi}
\usepackage{amsmath}
\usepackage[superscript]{cite}

\renewcommand{\vec}[1]{\mbox{\boldmath $#1$}}
\newcommand{\tensor}[1]{\mbox{\boldmath $#1$}}
\newcommand{\upd}{\mathrm{d}}

\makeatletter
\def\@citess#1{\textsuperscript{#1)}}
\makeatother

\makeatletter
\def\@biblabel#1{#1.}
\makeatother

\journal{Nihon Reoroji Gakkaishi (Journal of the Society of 
Rheology, Japan)}

\begin{document}

\begin{frontmatter}
\title{
Article:~\hfill~\\
Effects of pitched tips of novel kneading disks \\on melt mixing in twin-screw extrusion
}


\author[kyudai]{Yasuya Nakayama\corref{mycorrespondingauthor}
}
\cortext[mycorrespondingauthor]{Corresponding author}
\ead{nakayama@chem-eng.kyushu-u.ac.jp}

\author[kyudai]{Nariyoshi Nishihira}

\author[kyudai]{Toshihisa Kajiwara}

\author[jsw]{Hideki Tomiyama}
\author[jsw]{Takahide Takeuchi}
\author[jsw]{Koichi Kimura}

\address[kyudai]{Department of Chemical Engineering, Kyushu University,
Nishi-ku, Fukuoka 819-0395, Japan}
\address[jsw]{Hiroshima Plant, The Japan Steel Works Ltd. 1-6-1 Funakoshi-minami, Hiroshima 736-8602, Japan}

\begin{abstract}
In mixing highly viscous materials, like polymers, foods, and 
rubbers, the geometric structure of the mixing device is a 
determining factor for the quality of the mixing process.
In pitched-tip kneading disks~(ptKD), a novel type of mixing 
element, based on conventional kneading disks~(KD), the tip angle is 
modified to change the channel geometry as well as the drag 
ability of KD.
We discuss the effects of the tip angle in ptKD on mixing 
characteristics based on numerical simulation of the flow in the 
melt-mixing zone under different feed rates and a screw rotation 
speed.
It turns out that the passage of fluid through the high-stress 
regions increases in ptKD compared to conventional KD regardless 
of the directions and sizes of the tip angle, while the fluctuation in 
residence time stays at the same level as the conventional KD.
Furthermore, pitched tips of backward direction increase the mean 
applied stress on the fluid elements during its residence in 
the melt-mixing zone, suggesting the enhancement of dispersive mixing 
quality in ptKD.
These understandings of the role of the tip angle on KD can give a 
basic guide in selecting and designing suitable angle parameters 
of ptKD for different mixing purposes.
\end{abstract}


\end{frontmatter}


%
\section{Introduction}
Twin-screw devices are widely applied in various industries, 
including polymer processing, rubber compounding, food processing, 
and pharmaceutical development, because a high mixing 
performance can be achieved by the inter-meshing configuration of 
twin 
screws~\cite{Tadmor2006Principles,White2010Twin,2011Food,Rauwendaal2014Polymer,Kohlgruber2007CoRotating}.
In the plastics industry, twin-screw extruders are used to develop 
polymer materials with certain desired properties by mixing 
different fillers and additives with polymer melts.
For this melt-mixing process, several types of screw elements 
have been developed for different material 
processabilities and different mixing qualities, since the quality of the mixing process is primarily determined by the geometric 
structure of the mixing element. 
In selecting and/or developing a mixing element, one 
fundamental issue is a systematic understanding of the relation 
between the melt flow driven by the geometric structure of the mixing 
element and the mixing characteristics.

Among various types of mixing 
elements~\cite{Lawal1995Mechanisms,Lawal1995Simulation,Cheng1997Study,Bravo2000Numerical,Jaffer2000Experimental,Ishikawa20003D,Ishikawa20013D,Ishikawa2002Flow,Funatsu20023D,Bravo2004Study,Ishikawa2006Tipclearance,Alsteens2004Parametric,Malik20053D,Kalyon2007Integrated,Kohlgruber2007CoRotating,Zhang2009Numerical,Hirata2013Experimental,Hirata2014Effectiveness,Nakayama2016Strain}, 
kneading disks~(KD) are the most commonly used due to their mixing 
efficiency.
In order to tune the mixing characteristics of conventional KD 
for application to various purposes, a modified type of KD has 
been proposed, called ``pitched-tip kneading 
disks''~(ptKD)~\cite{Nakayama2011Meltmixing}.
In ptKD, two disk tips are pitched to the screw axes, whereas the 
tips in the conventional KD are parallel to the screw axes.
The combined choice of the tip angle and the disk-stagger angle can make it 
possible to control both the distributive and dispersive 
mixing abilities better than in conventional KD.
In the previous study~\cite{Nakayama2011Meltmixing}, it has been 
found that different combinations of tip angle and disk-stagger 
angle actually lead to different mixing characteristics.
Nonetheless, the essential role of the tip angle, specifically its 
direction and size, on the modification of the flow pattern and the 
resulting mixing characteristics are still unclear.

In this article, we investigate how pitched tips modify the 
mixing characteristics of conventional KD in twin-screw extrusion. 
For this purpose, we consider different directions and sizes of 
the tip angle on a neutrally staggered KD, and discuss their mixing 
characteristics using numerical simulation of 
the polymer-melt flow in the kneading zone.

\section{Pitched-tip kneading elements}
\begin{figure}
 \centering
 \includegraphics[width=.5\hsize]{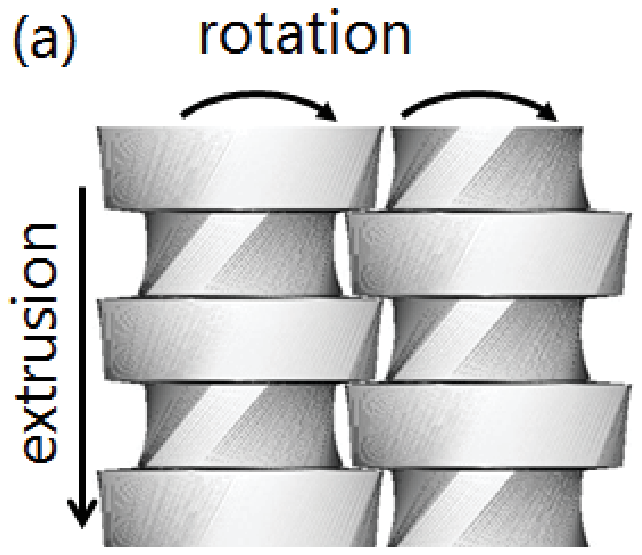} 
\\
\vspace*{2ex}
 \includegraphics[width=.5\hsize]{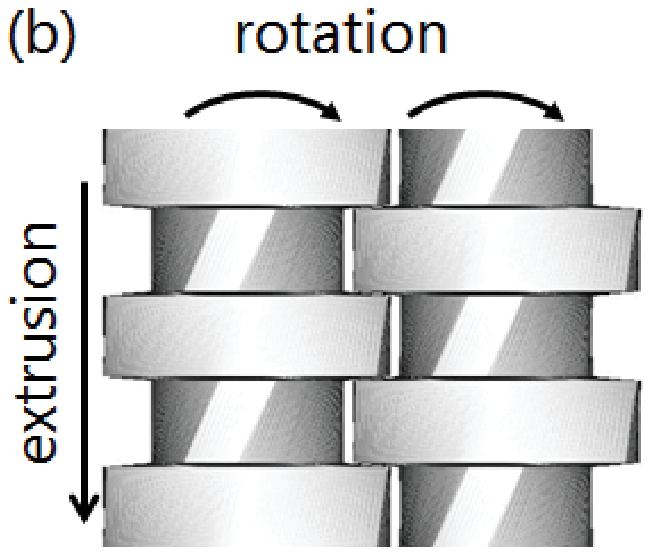}
\\
\vspace*{2ex}
 \includegraphics[width=.5\hsize]{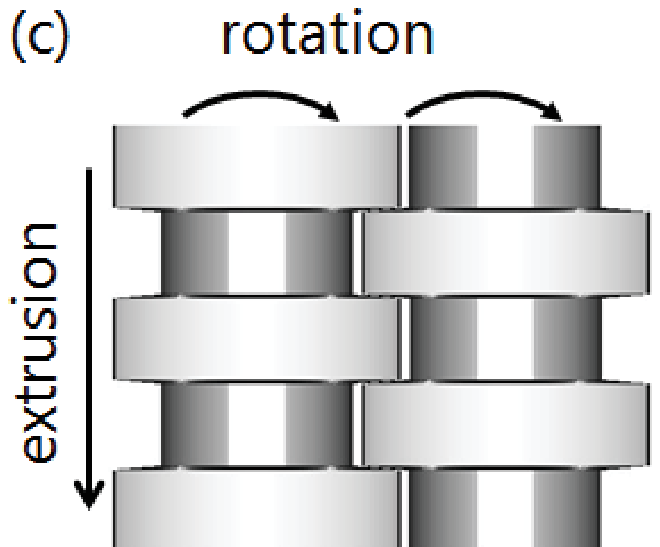}
\\
\vspace*{2ex}
 \includegraphics[width=.5\hsize]{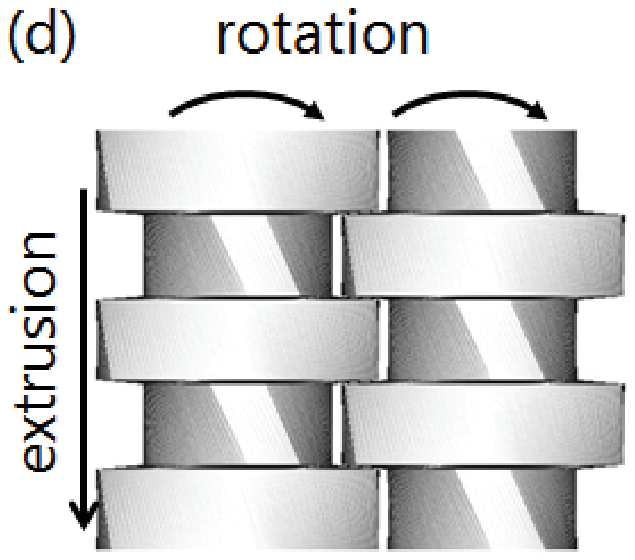}
\\
\vspace*{2ex}
\includegraphics[width=.5\hsize]{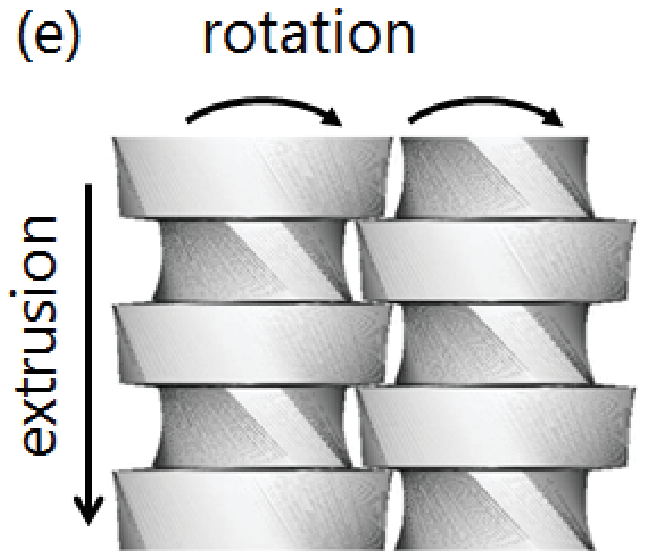}
\caption{\texttt{Top view of pitched-tip kneading disks used in this study. 
Stagger angle of adjacent blocks is set to a value of 90\(^{\circ}\) for all the kneading disks
while tip angle varies from -30\(^{\circ}\) to 30\(^{\circ}\).
(a) forwarding ptKD with a tip angle of 30\(^{\circ}\)~(Ft30), 
(b) forwarding ptKD with a tip angle of 15\(^{\circ}\)~(Ft15), 
(c) conventional KD with a tip angle of 0\(^{\circ}\)~(Nt), 
(d) backwarding ptKD with a tip angle of -15\(^{\circ}\)~(Bt15), 
and
(e) backwarding ptKD with a tip angle of -30\(^{\circ}\)~(Bt30).
}}
\label{fig1}
\end{figure}

Conventional kneading disks~(KD) are composed of several oval blocks 
combined with a stagger angle between two adjacent blocks~(Fig.~\ref{fig1}(c)).
The tips of the blocks used in conventional KD are parallel to the screw axes.
In contrast to this, in pitched-tip KD, the tips of the blocks are 
twisted to the screw axes~\cite{Nakayama2011Meltmixing}.
Pitched tips modify the channel geometry of the conventional KD. 
Furthermore, pitched tips add additional drag ability along with 
screw rotation to conventional KD.
The tip angles associated with forward pump will be called ``forward
tip''~(Ft) and define its tip angle as positive. 
Conversely, for negative tip angles, pitched tips add backward 
pump to a block, so that we will call it ``backward tip''~(Bt).
Along with this terminology, we will call the conventional block with zero tip 
angle the ``neutral tip''~(Nt).

The typical ptKD has been designed by adjusting both the tip angle and 
the disk-stagger angle, and therefore its mixing characteristics are 
determined by the combination of the tip angle and the disk-stagger angle~\cite{Nakayama2011Meltmixing}.
Effects solely due to the pitched tip have not been discussed 
systematically in previous works.
In this paper, we investigate the effects of the pitched tip 
direction and its angle on the mixing characteristics. For this 
purpose, ptKDs with a disk-stagger angle and different tip angles,
shown in Fig.~\ref{fig1}, are chosen. Based on five-block kneading 
elements of \(L/D=1.47\) with a disk-stagger angle of 
90\(^{\circ}\), pitched tips with different tip angles  from 
-30\(^{\circ}\) to 30\(^{\circ}\) are arranged.

\section{Numerical Simulation}
The flow of a polymer melt in the melt-mixing zone of a twin-screw extruder
has been numerically solved to allow understanding the relation between
the melt-mixing characteristics and 
the geometry of the kneading elements.
The numerical method to study the melt-mixing zone in this paper follows that 
of the previous paper~\cite{Nakayama2011Meltmixing}. Here we 
give a summary of the numerical simulation.
We focus on the situation where the material fully fills the channel.
The Reynolds number is assumed to be much less than unity, so that
inertial effects are neglected.
The flow is assumed to be incompressible, and in a pseudo-steady state to
screw rotation, as has often been assumed in polymer flow in twin-screw
extruders~\cite{Ishikawa20003D,Bravo2004Study,Malik20053D,Zhang2009Numerical,Nakayama2011Meltmixing,SarhangiFard2013Simulation,Rathod2013Effect,Hirata2014Effectiveness}.
With these assumptions, the governing equations become
\begin{align}
 \vec{\nabla}\cdot\vec{v}&=0,
\label{eq:incompressibility}
\\
\vec{0} &= -\vec{\nabla}p +\vec{\nabla}\cdot\tensor{\tau},
\label{eq:stokes_equation}
\\
\rho C_{p}\vec{v}\cdot\vec{\nabla}T &= k \nabla^{2}T
 +\tensor{\tau}:\tensor{D},
\label{eq:temperature_equation}
\end{align}
where \(\vec{v}\) is the velocity,  
\(p\) is the pressure, \(\tensor{\tau}\) is the deviatoric 
stress, \(\rho\) is the mass density, \(c_{p}\) is the specific 
heat capacity, \(T\) is the temperature, \(k\) is the thermal 
conductivity, 
and \(\tensor{D}=\left[\vec{\nabla}\vec{v}+\left(\vec{\nabla}\vec
{v}\right)^{T}\right]/2\) is the strain-rate tensor, where \((.)^{T}\) indicates the transpose.

The fluid is assumed to be a viscous shear-thinning fluid that follows
Cross--exponential viscosity~\cite{Cross1965Rheology}, 
\begin{align}
\tensor{\tau} &= 2\eta \tensor{D},
\label{eq:viscous_stress}
\\
 \eta(\dot{\gamma}, T)&=\frac{\eta_{0}(T_{0})H(T,T_{0})}{1+\left(
\lambda(T_{0}) H(T,T_{0})\dot{
\gamma}
\right)^{1-n}},
\label{eq:cross_model}
\\
H(T,T_{0}) &=\exp\left[-\beta(T-T_{0})\right],
\label{eq:exponential_model}
\\
\dot{\gamma} &= \sqrt{2\tensor{D}:\tensor{D}},
\end{align}
whose parameters are
obtained by fitting the shear viscosity of a polypropylene melt taken
from~\cite{Ishikawa20003D}, and the values are 
\(T_{0}=473.15\)\,K,
\(\eta_{0}(T_{0})=32783\)\,Pa\(\cdot\)s,
\(n=0.33\),
\(\beta=0.0208\)\,K\(^{-1}\),
and
\(\lambda(T_{0})=1.702\)\,s.
The mass density, specific heat capacity, and thermal conductivity are
taken from~\cite{Ishikawa20003D} as well, and the values are
\(\rho=735.0\)\;kg/m\(^{3}\), \(c_{p}=2100\)\;J/(kg\(\cdot\)K), and
\(k=0.15\)\;W/(m\(\cdot\)K).  

As operational conditions, the screw rotation speed is set to 
200\,rpm, while the volume flow rate varies in the range of 
5--120\,cm\(^{3}\)/s (\(\approx\)13--318\;kg/h), so that the 
value of \(Q/N\) lies in the range of 1.5--36\,cm\(^{3}\).
The no-slip condition on the velocity at the barrel and screw
surfaces is assumed. 
The velocity at the inlet and outlet boundaries was set to be 
uniform under the given volumetric flow rate.
The pressure at the outlet boundary was fixed to be a constant value.
The temperatures on the barrel surface and at the inlet boundary were
set to 473.15~K and 453.15~K, respectively. The natural boundary conditions
for the temperature equation in the exit boundary plane and the screw
surface were assigned.
The time evolution of the velocity and temperature fields was constructed
with the converged fields for every three degrees of screw rotation.
The trajectories of the passive tracers were determined based on 
the solved velocity field.  The set of tracer trajectories was 
utilized to compute the residence time 
distribution~\cite{Levenspiel1998Chemical,Kohlgruber2007CoRotating}
 and 
to investigate the flow history and mixing process.
The Lagrangian-history average of a quantity \(f\) over the trajectory of
\(\alpha\)th tracer is defined as
\begin{align}
 \overline{f_{\alpha}}^{T_{\alpha}} &=\frac{1}{T_{\alpha}}\int_{0}^{T_{\alpha}}\upd
 s\int\upd\vec{x}\delta\left(
\vec{x}-\vec{X}_{\alpha}(s)
\right)f(\vec{x},s),
\label{eq:lagrangian_average_of_f}
\end{align}
where \(T_{\alpha}\) and \(\vec{X}_{\alpha}(.)\) are the residence time
and position of the \(\alpha\)th tracer, respectively, and \(\delta(.)\)
is the Dirac delta distribution.
The statistical distribution of \(\overline{f_{\alpha}}^{T_{\alpha}}\)
characterizes the global flow characteristics in the melt-mixing zone.

Initially, 2000 points were uniformly distributed in a certain section
relative to the axial position, which was arbitrarily set in the second disk.
They were advected until they reached the outlet section.
When computing the tracer advection, some tracers went out of bounds
because the time resolution of the velocity field was limited.  To
circumvent the effect of the lost tracers on the statistics, we set the
number of initial points to 2000, which ensured that a sufficient number
of points reached the outlet. 
\section{Results and Discussion}
\begin{figure}[htbp]
 \centering
 \includegraphics[width=.8\hsize]{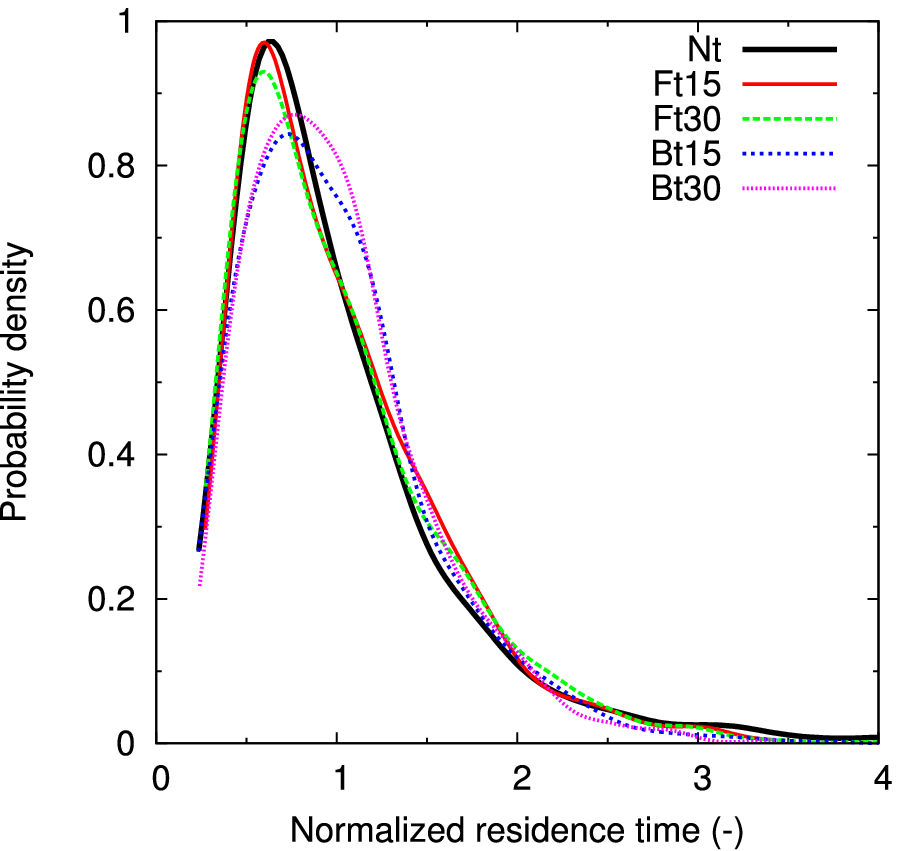}
\caption{\texttt{Normalized residence time distribution for Ft30, Ft15, Nt, Bt15, and Bt30 
under \(Q/N=9\)\,cm\(^{3}\). Residence time is normalized by the mean residence time.
}}
\label{fig2}
\end{figure}

\begin{figure}[htbp]
 \centering
 \includegraphics[width=.8\hsize]{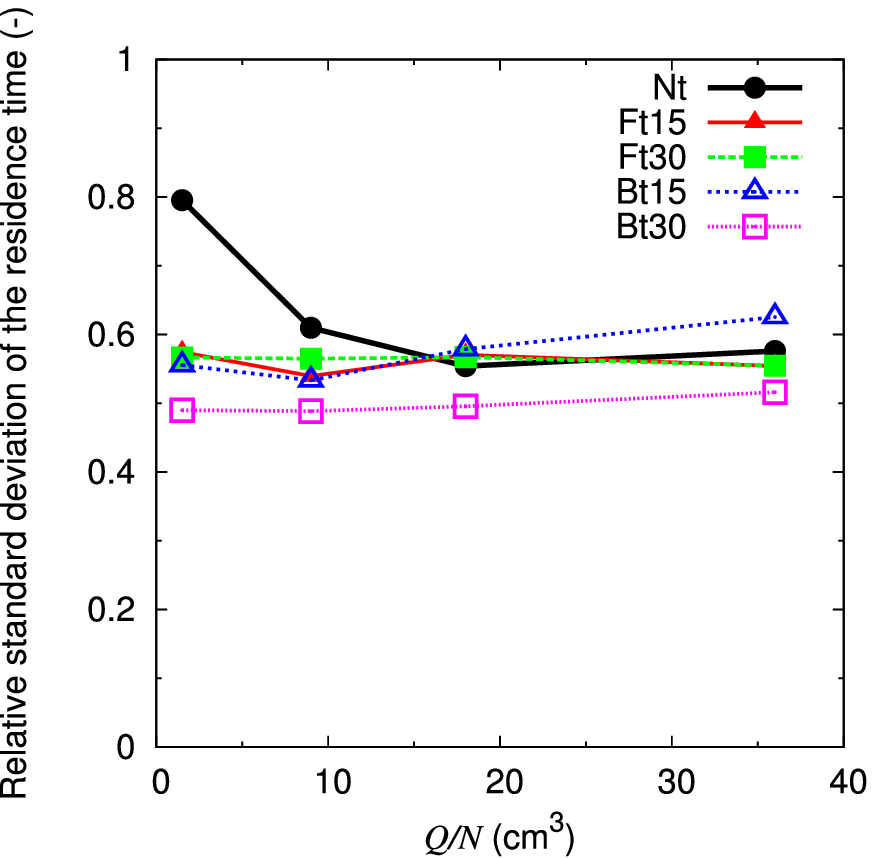}
\caption{\texttt{
Relative standard deviation of the 
residence time to its mean value as a function of \(Q/N\) for Ft30, Ft15, Nt, Bt15, and Bt30.}}
\label{fig3}
\end{figure}

In order to see the effect of the tip angle on the inhomogeneity of 
cross-sectional mixing, we computed the fluctuation of the residence time. 
Figure~\ref{fig2} shows the probability density of the residence 
time normalized by the mean residence time for five different 
ptKDs under \(Q/N=9\)\,cm\(^{3}\).
From this figure, we observe that 
the residence time distribution is almost unaffected by the forward 
tip angle while the fraction of the longer residence 
increases with a backward tip angle.
In Fig.~\ref{fig3}, there is drawn the relative standard deviation of the 
residence time to its mean value as a function of \(Q/N\).
It turns out that the fluctuation of the residence time in ptKDs 
hardly depends on \(Q/N\), and takes the value of 0.5--0.6.
At higher values of \(Q/N\), the residence time fluctuation is 
approximately the same level as that of the conventional KD.
This fact suggests that ptKDs have as good mixing 
ability as the conventional KD with the same disk-stagger angle 
at sufficiently high \(Q/N\).


The effect of the tip angle on the residence time distribution 
depends on the tip direction.
While for Ft type the residence time fluctuation does not show 
any angle-dependence, for Bt type the residence time fluctuation 
decreases slightly with the backward tip angle. Although this slight 
change of residence time fluctuation reflects the effect of the tip angle,
this change does not seem to be substantial in terms of mixing ability.
In short, the residence time distribution is not substantially modified by 
the addition of a tip angle to a conventional KD.

\begin{figure}[htbp]
 \centering
 \includegraphics[width=.8\hsize]{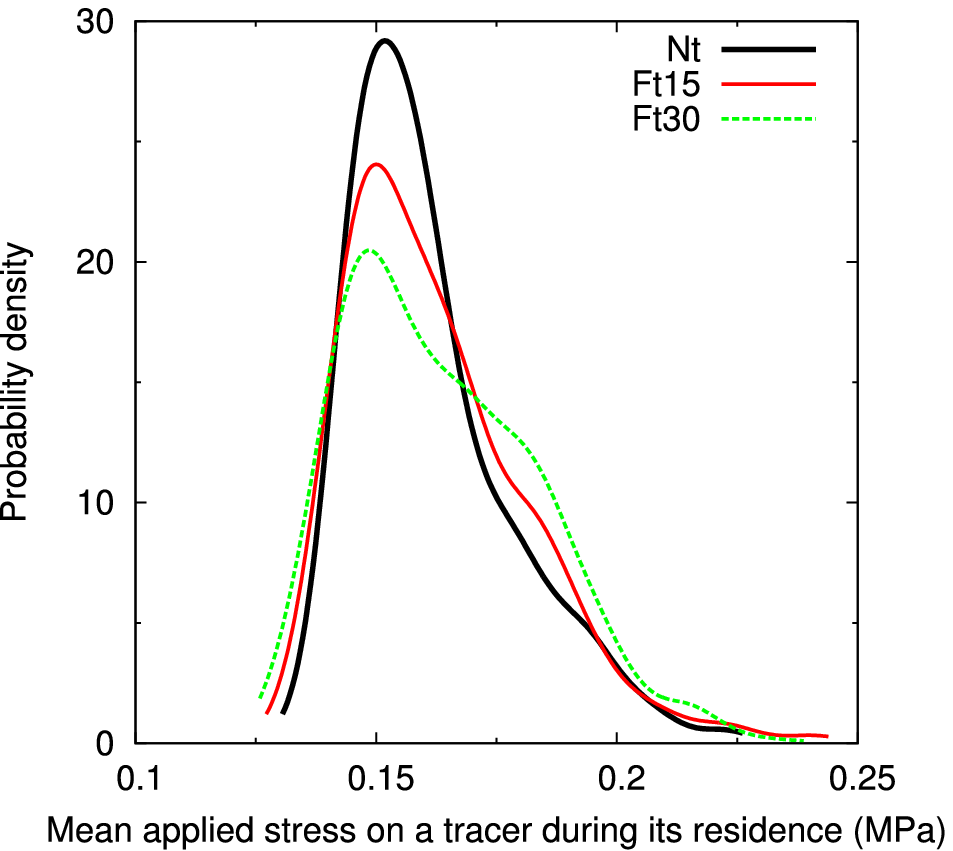}
\caption{\texttt{Probability density function of the applied stress 
during residence, \(\overline{\sigma_{\alpha}}^{T_{\alpha}}\), 
for Ft30, Ft15, and Nt under \(Q/N=9\)\,cm\(^{3}\).}}
\label{fig4}
\end{figure}
\begin{figure}[htbp]
 \centering
 \includegraphics[width=.75\hsize]{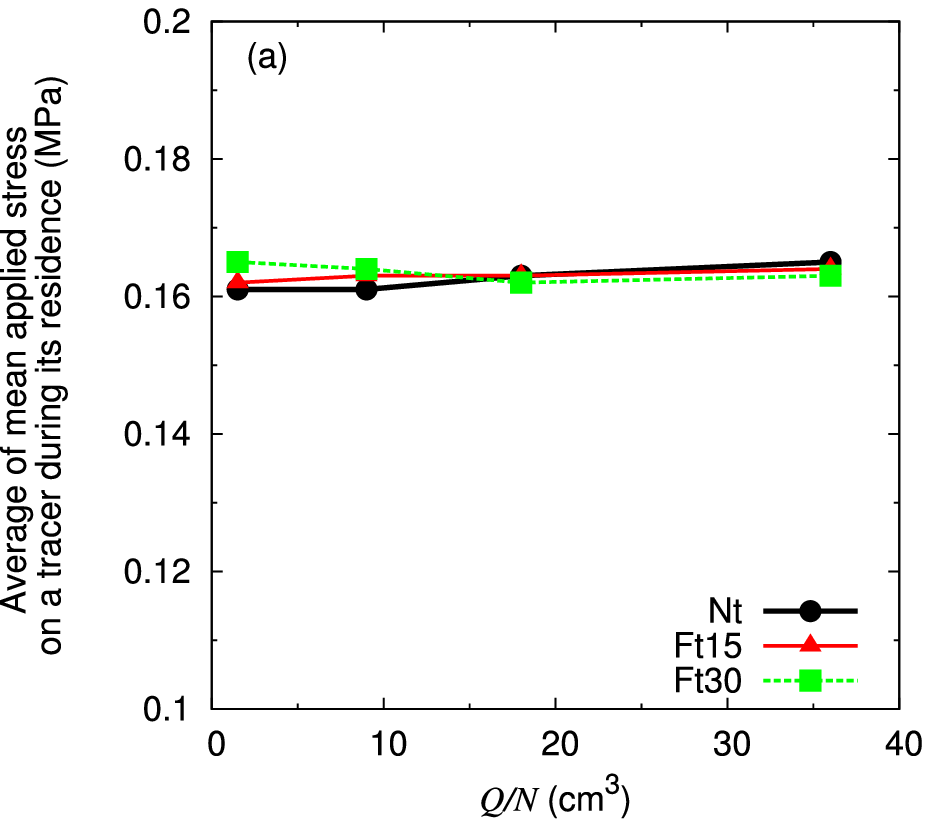}
 \includegraphics[width=.75\hsize]{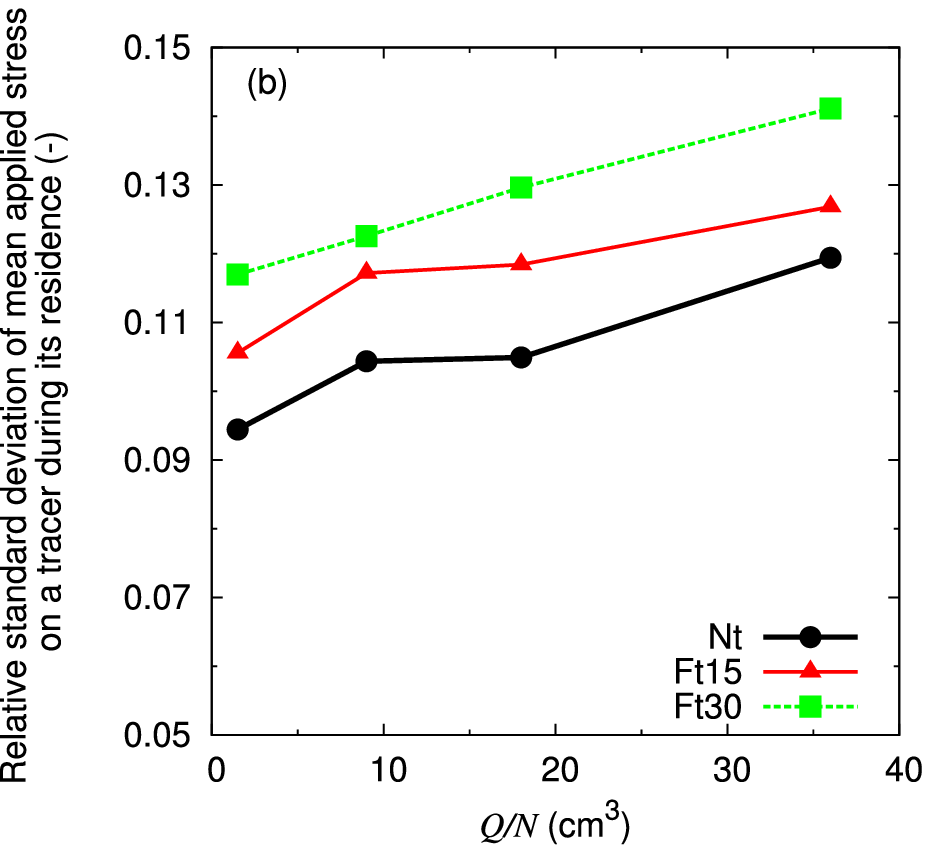}
\caption{\texttt{
(a) Average value and (b) the relative standard deviation of  
\(\overline{\sigma_{\alpha}}^{T_{\alpha}}\) 
as a function of \(Q/N\) for Ft30, Ft15, and Nt.
}}
\label{fig5}
\end{figure}
Next, we discuss the effects of a pitched tip on dispersive mixing.
The mean applied stress during residence, 
\(\overline{\sigma_{\alpha}}^{T_{\alpha}}\), is calculated by 
inserting the stress magnitude 
\(\sigma=\sqrt{(3/2)\tensor{\tau}:\tensor{\tau}}\) into 
Eq.~(\ref{eq:lagrangian_average_of_f}).
Figure~\ref{fig4} shows the probability density function of 
\(\overline{\sigma_{\alpha}}^{T_{\alpha}}\)
for Ft15, Ft30, and Nt under \(Q/N=9\)\,cm\(^{3}\).
As the forward tip angle increases, the fluctuation of 
\(\overline{\sigma_{\alpha}}^{T_{\alpha}}\) increases, whereas the 
average of \(\overline{\sigma_{\alpha}}^{T_{\alpha}}\) stays at a 
similar level as that of Nt.
The average value and the relative standard deviation of  
\(\overline{\sigma_{\alpha}}^{T_{\alpha}}\) 
as a function of \(Q/N\) are drawn in Figs.~\ref{fig5}(a) 
and (b), respectively.
The average value of \(\overline{\sigma_{\alpha}}^{T_{\alpha}}\) 
is insensitive to \(Q/N\), and takes a similar value as that 
of Nt.
In contrast, the fluctuation of 
\(\overline{\sigma_{\alpha}}^{T_{\alpha}}\) increases with \(Q/N\),
and is larger for a larger forward tip angle.
These results clearly indicate that the forward tip increases the 
inhomogeneity of dispersive mixing quality especially at high 
throughput operation.
%

\begin{figure}[htbp]
 \centering
 \includegraphics[width=.8\hsize]{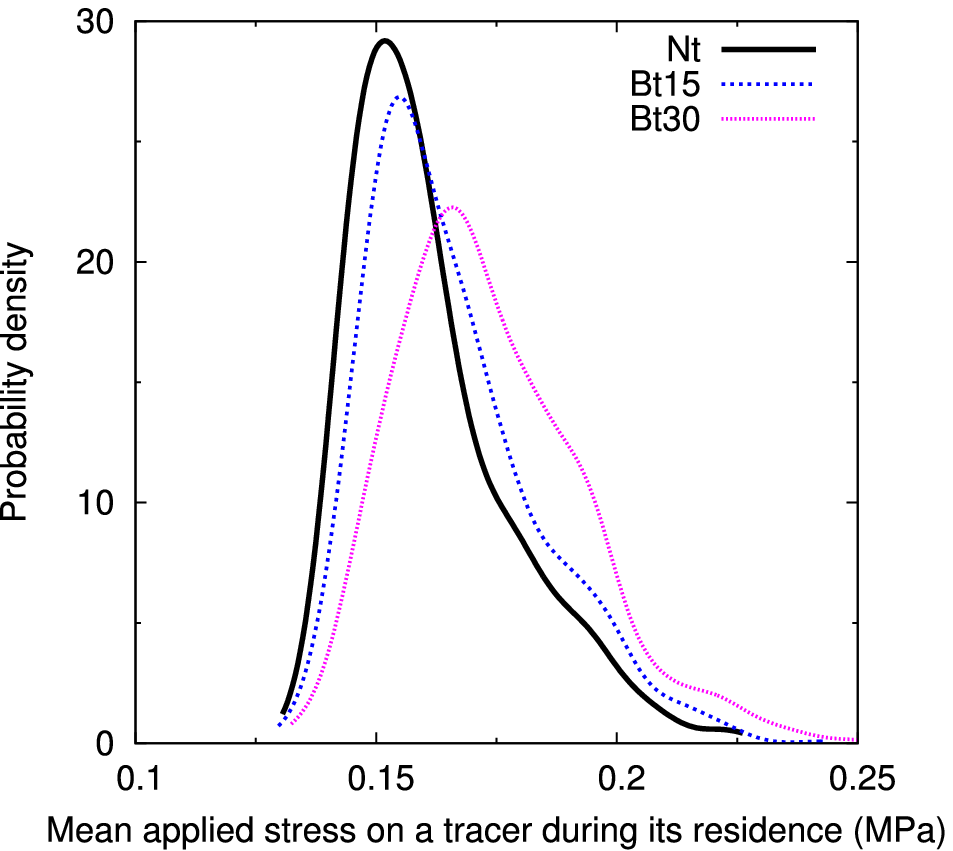}
\caption{
\texttt{Probability density function of the applied stress 
during residence, \(\overline{\sigma_{\alpha}}^{T_{\alpha}}\), 
for Bt30, Bt15, and Nt under \(Q/N=9\)\,cm\(^{3}\).}}
\label{fig6}
\end{figure}

\begin{figure}[htbp]
 \centering
 \includegraphics[width=.75\hsize]{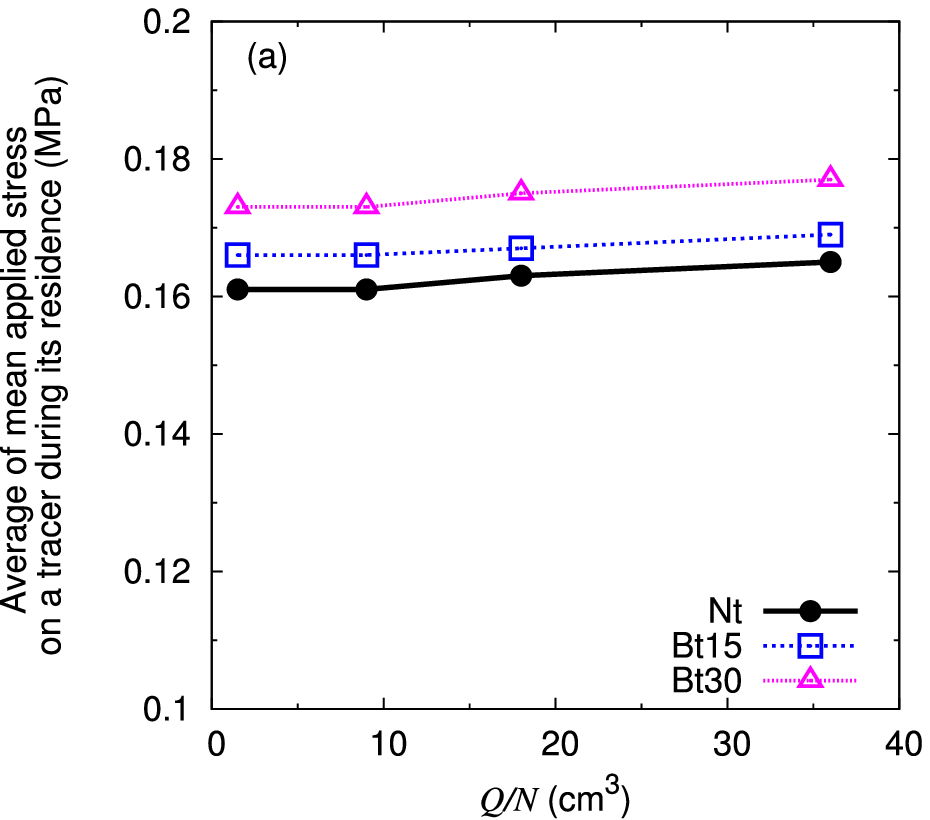}
 \includegraphics[width=.75\hsize]{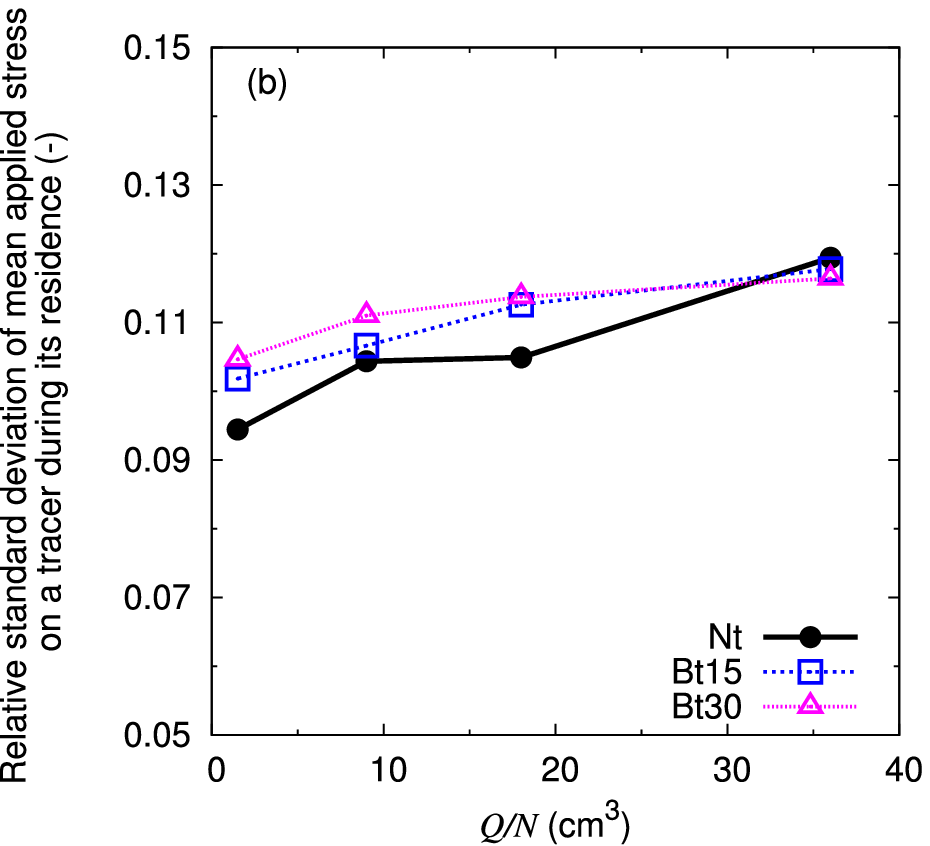}
\caption{\texttt{
(a) Average value and (b) the relative standard deviation of  
\(\overline{\sigma_{\alpha}}^{T_{\alpha}}\) 
as a function of \(Q/N\) for Bt30, Bt15, and Nt.
}}
\label{fig7}
\end{figure}

Figure~\ref{fig6} shows the probability density function of 
\(\overline{\sigma_{\alpha}}^{T_{\alpha}}\)
for Bt15, Bt30, and Nt under \(Q/N=9\)\,cm\(^{3}\).
In Fig.~\ref{fig6}, we observe an increase both in the average level and 
the fluctuation of \(\overline{\sigma_{\alpha}}^{T_{\alpha}}\) 
with an increase in the backward tip angle.
The average value and the relative standard deviation of  
\(\overline{\sigma_{\alpha}}^{T_{\alpha}}\) 
as a function of \(Q/N\) are drawn in Figs.~\ref{fig7}(a) and (b), respectively.
In Fig.~\ref{fig7},
the average value of \(\overline{\sigma_{\alpha}}^{T_{\alpha}}\) 
becomes larger for a larger backward tip angle, suggesting that the Bt type 
enhances the dispersive mixing ability compared to the Nt type.
In contrast, the fluctuation of 
\(\overline{\sigma_{\alpha}}^{T_{\alpha}}\) is rather insensitive 
to the backward tip angle. That is, irrespective of Bt angle,
the inhomogeneity of the dispersive mixing is at the same level 
as that of the Nt type 
With an increase of \(Q/N\), the fluctuation of 
\(\overline{\sigma_{\alpha}}^{T_{\alpha}}\) for Bt slightly 
increases, suggesting that inhomogeneity in the dispersive 
mixing is smaller for lower \(Q/N\).
In short, for Bt type, both high level and small inhomogeneity of 
\(\overline{\sigma_{\alpha}}^{T_{\alpha}}\) is achieved at low 
\(Q/N\) conditions, indicating that that dispersive mixing in Bt 
type is most effective when \(Q/N\) is low and the Bt angle is 
large.


In order to further study the effects of the tip-angle on dispersive mixing,
the passage of fluid at high-stress regions has been investigated.
Irrespective of the tip angle, 
the highest shear stress is achieved in the small gap regions like 
the tip--barrel clearance and the inter-meshing region.
The substantial dispersion process mainly takes place when fluid elements 
pass through such high-stress regions. 
However, since the high-stress region occupies only a small 
fraction of the whole channel, the fraction of the fluid passing 
through these regions and the number of passages are essentially 
determined by the flow pattern at the low-stress (low-shear-rate) 
regions.
Thus, the geometric structures of the mixing elements are directly 
responsible for the passage of fluid through the high-stress regions, 
eventually leading to a better dispersive mixing performance.

\begin{figure}[htbp]
 \centering
 \includegraphics[width=.75\hsize]{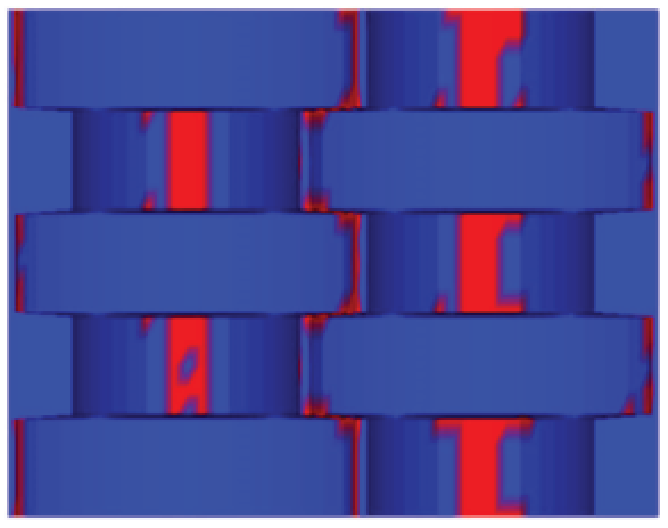}
\caption{\texttt{
High-stress region for Nt type under \(Q/N=18\)\;cm\(^{3}\) is  
drawn in red. In this case, the threshold value of the shear 
stress magnitude,  \(\sigma\), is arbitrarily set to 0.35\;MPa. 
}}
\label{figa}
\end{figure}
\begin{figure}[htbp]
 \centering
 \includegraphics[width=.75\hsize]{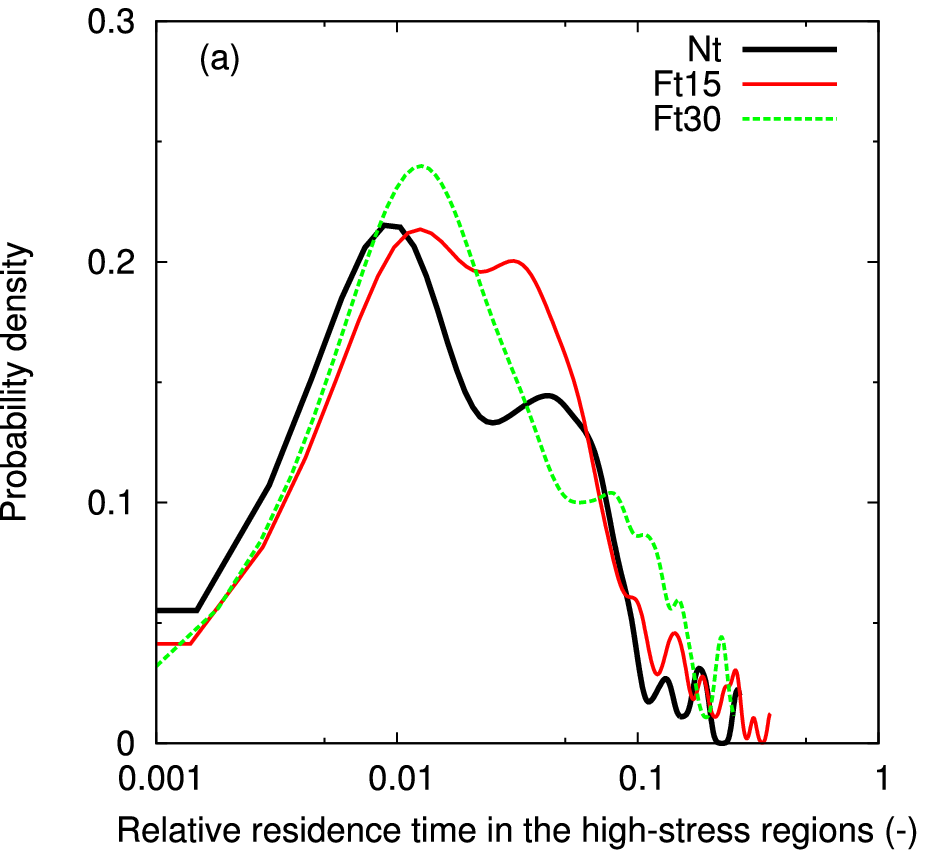}
 \includegraphics[width=.75\hsize]{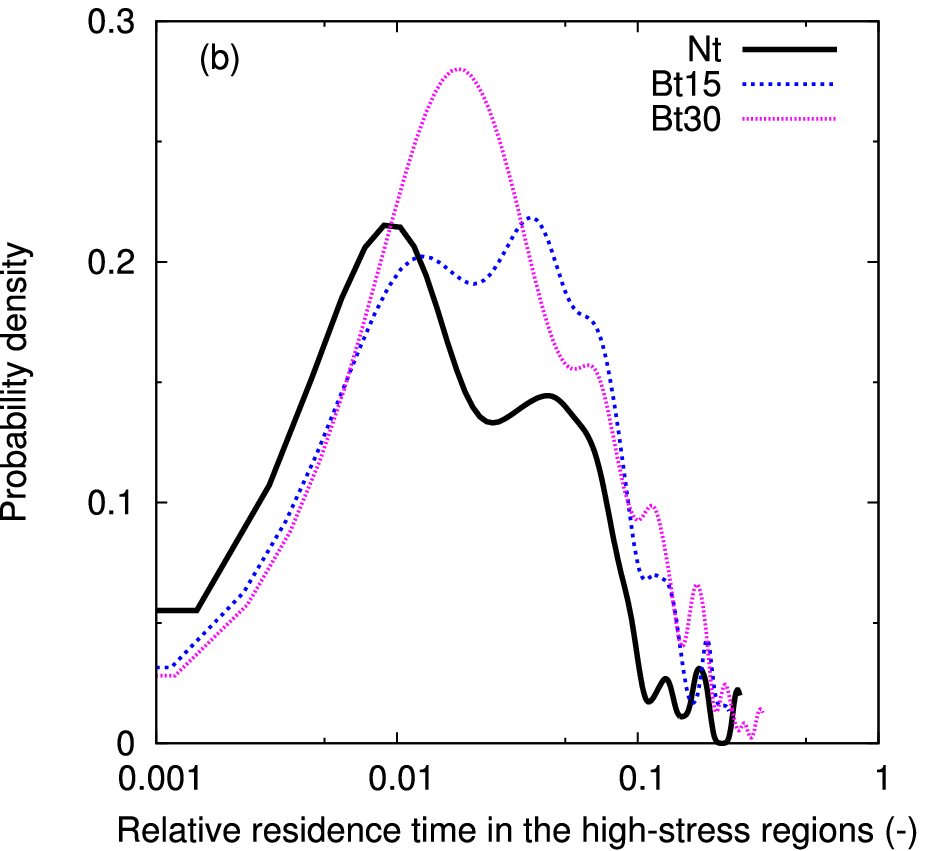}
\caption{\texttt{
Probability density function of the ratio of the residence time in the high-stress regions to the overall residence time, 
\(t_{h,\alpha}/T_{\alpha}\), under \(Q/N=9\)\,cm\(^{3}\): 
(a) for Ft30, Ft15, and Nt, and (b) for Bt30, Bt15, and Nt.
}}
\label{fig8}
\end{figure}
\begin{figure}[htbp]
 \centering
 \includegraphics[width=.75\hsize]{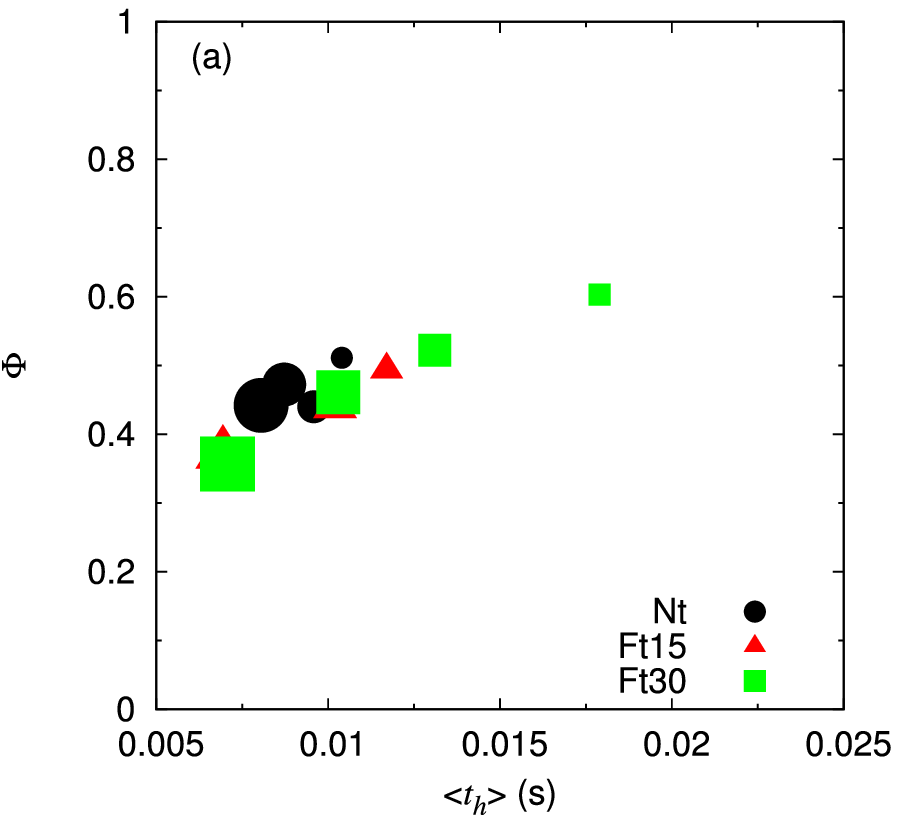}
 \includegraphics[width=.75\hsize]{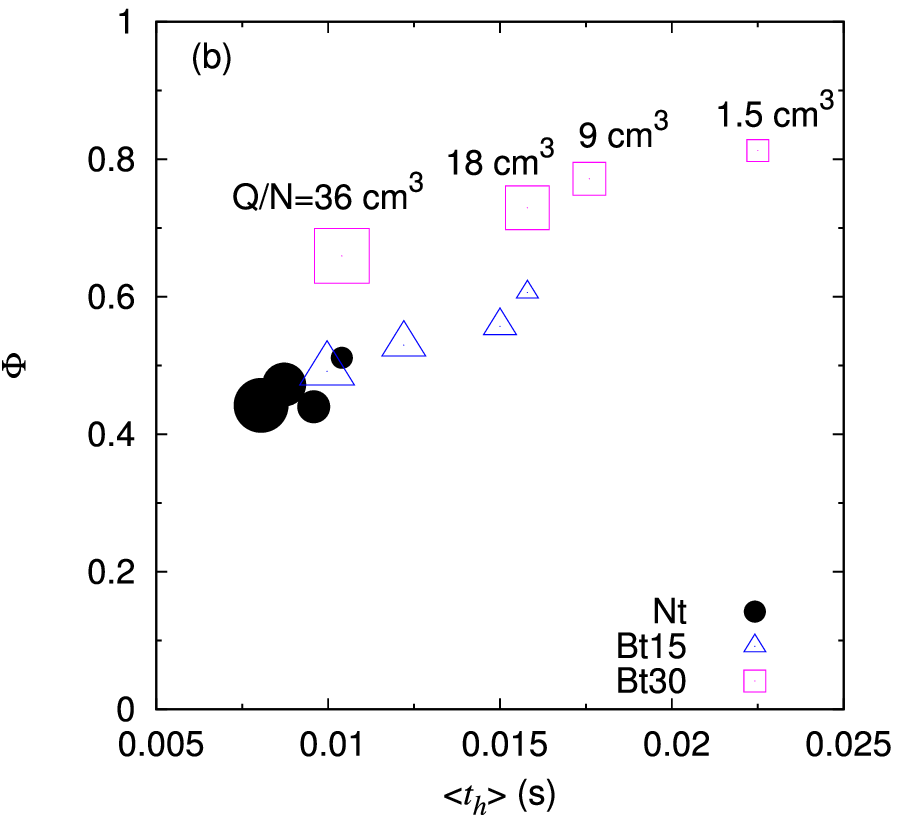}
\caption{%
\texttt{Fraction of the tracers passing through the high-stress 
regions, \(\Phi\), and the mean residence time in those regions, 
\(\langle t_{h}\rangle\), 
for different values of \(Q/N=\)1.5, 9, 
 18, and 36\;cm\(^{3}\): (a) 
for Ft30, Ft15, and Nt, and (b) for Bt30, Bt15, and Nt.
The larger symbol is for the larger value of \(Q/N\).}}
\label{fig9}
\end{figure}

By setting a threshold value for the magnitude of the shear 
stress tensor, the regions of tip--barrel clearance and the screw 
inter-meshing region are defined as the high-stress regions. 
For instance, the region where the shear stress magnitude is 
larger than 0.35\;MPa for Nt type under \(Q/N=18\)\;cm\(^{3}\) is 
shown in Fig.~\ref{figa}, which clearly shows that the high-stress 
region is localized at around tip-clearance and inter-meshing regions.
We calculated the fraction of the tracers passing through the high-stress 
regions as well as the residence time in those regions.
For each tracer, the residence time in the high-stress regions, \(t_{h,\alpha}\), 
as well as the residence time in the whole mixing zone, \(T_{\alpha}\), has been studied.
Figures~\ref{fig8}(a) and (b) show the probability density 
function of the ratio \(t_{h,\alpha}/T_{\alpha}\) under \(Q/N=9\)\,cm\(^{3}\).
At a first look at Figs.~\ref{fig8}(a) and (b), the residence 
time in the high-stress regions is at most 0.1 during the residence 
in whole zone, irrespective of the tip angle.
Both for Ft and Bt types, the residence in the high-stress 
regions becomes longer compared to the Nt type, indicating that 
the pitched-tip geometry can enhance the efficiency of conveying the 
fluid to the high-stress regions.
The increase of the relative residence in the high-stress regions 
is more pronounced with the Bt type.
This observation partly explains the enhancement of the 
dispersive mixing by the pitched tip.
From Figs.~\ref{fig8}(a) and (b), we also observe that 
the probability of vanishing \(t_{h,\alpha}\) takes a finite value, 
indicating that not all the fluid elements pass through the 
high-stress regions. 
In order to study this aspect, we discuss the fraction of the 
tracers passing the high-stress regions, denoted by \(\Phi\).
In Figs.~\ref{fig9}(a) and (b), \(\Phi\) is plotted to the mean 
residence time of high-stress regions, \(\langle t_{h}\rangle\), 
for different \(Q/N\) values, in which the larger symbol is for 
the larger value of \(Q/N\).
In the plot of \(\Phi_{h}\) and \(\langle t_{h}\rangle \), 
the upper locations means that a large fraction of tracers pass 
through the high-stress regions, while the right locations means 
that the tracers stay long time in the high-stress regions. Hence, 
the upper and right locations indicate a high potential for 
dispersive mixing.

For conventional Nt type, a fraction, 0.4--0.5, of the fluid passes through the high-stress region for the \(Q/N\) value studied.
Furthermore, \(\Phi_{h}\) and \(\langle t_{h}\rangle \) are rather insensitive to the value of \(Q/N\).
In contrast, for Ft and Bt types, 
\(\Phi_{h}\) and \(\langle t_{h}\rangle \) show a relatively large dependence on \(Q/N\).
We observe a general trend such that,
as \(Q/N\) decreases, both \(\Phi_{h}\) and \(\langle t_{h}\rangle \) increase.
In addition, the characteristics of the passage through the high-stress 
region have different tip-angle dependencies with different directions of 
the tip angle.
While for Ft type, \(\Phi_{h}\) and \(\langle t_{h}\rangle \) 
does not depend on the tip angle, for Bt type, \(\Phi_{h}\) and 
\(\langle t_{h}\rangle \) take larger values for larger backward 
tip angles.
This fact is consistent with the tip-angle dependence of the mean 
applied stress during residence in Fig.~\ref{fig5}(a) and \ref{fig7}(a).
The characteristics of the passage through the high-stress region reveal that
the backward tip is effective at enhancing the dispersive mixing ability. 
From the data above, it turns out that the pitched tip modifies the 
dispersive mixing ability of kneading disks, and the backward tip 
enhanced the potential of dispersive mixing while keeping the 
distributive mixing ability.
We discuss the relation between the flow pattern induced by the 
geometry of the pitched-tip KD and the dispersive mixing ability.
Pitched tips add an additional drag ability along the screw rotation to conventional KD.
For backward tips increase the drag in the backward extrusion 
direction, for \(Q>0\) condition, the pressure drop in Bt type 
becomes larger than that in Nt, leading to  the pressure flow in 
the extrusion direction being more pronounced in Bt.
The enhanced pressure flow in Bt promotes the passage of fluid across 
the small gap at the backward tips, resulting in an increase of the 
passage through the high-stress regions and the mean applied stress.
In other words, the geometry of backward tips cause a 
counteracting effect by blocking against the forward flow and the 
additional forward pressure flow while keeping the leakage flow 
in the inter-disk spaces. 
Since this effect is driven by the screw rotation, it is 
maximized with the small \(Q/N\) condition, which is clearly 
demonstrated by the characteristics of the passage through the 
high-stress region in Fig.~\ref{fig9}.

In the case of Ft type, since the forward pitched tips increase the 
drag in the forward direction,  for \(Q>0\) conditions the pressure 
drop become smaller than for Nt, or even becomes negative,
depending on \(Q/N\).
Thus, the counteracting effect of the pressure flow and the 
blockage by the tips is weaker in the Ft type than in the Bt type.

In short, the geometric modification of conventional KD by pitched tips 
induces both a change in the channel geometry 
and a modification of the pressure flow, causing a change in 
dispersive mixing ability.

\section{Conclusions}
In pitched-tip kneading disks~(ptKD), a novel melt-mixing element 
used in twin-screw extrusion, the effects of the direction and size 
of the tip angle on the mixing characteristics were investigated based on 
the numerical simulation of a three-dimensional flow.
Under a material feed rate and a screw rotation speed, the tip angle 
on a conventional kneading-disk~(KD) element modifies the following two 
properties of the flow: (i) additional drag ability caused by screw 
rotation modifies the pressure flow, (ii) the pitched-tip acts as a 
blockage to the pressure flow induced by the pitched-tip.
These counteracting effects result in an increase of the fraction 
of the fluid elements passing through the high-stress regions, 
leading to a modification of the dispersive mixing ability.

Forward tips on neutrally staggered KD increase the inhomogeneity of 
the applied stress during residence compared to the conventional KD,
while keeping the residence time fluctuation.
In contrast, backward tips on neutrally staggered KD increase 
the mean applied stress during residence, suggesting an enhancement of 
the dispersive mixing ability. 
These effects are more pronounced for larger tip angles, and are 
dependent on the ratio of the material feed rate to the screw 
rotation speed.
These findings about the role of the tip angle can be useful for 
a better understanding of the mixing characteristics of a general class 
of ptKD elements and for choosing the optimal combination of the 
disk-stagger angle and tip angle.
%


\bigskip

\section*{Acknowledgments}
The numerical calculations have been partly carried out using the
computer facilities at the Research Institute for Information Technology
at Kyushu University.
This work has been supported by Grants-in-Aid for Scientific Research
(JSPS KAKENHI) under Grant Nos.~26400433, 24656473, and 15H04175.
\renewcommand{\refname} {REFERENCES}

\end{document}